\documentclass[journal]{IEEEtran}
\ifCLASSINFOpdf
\else
\fi
\usepackage{graphicx}
\usepackage{float}
\begin{document}

\title{Implementation of True Random Number Generator based on Double-Scroll Attractor circuit with GST memristor emulator}

\author{Togzhan~Abzhanova, 
    Irina Dolzhikova,~\IEEEmembership{ Student Member,~IEEE,}
Alex Pappachen James,~\IEEEmembership{Senior Member,~IEEE,}
        Electrical and Computer Engineering Department, Nazarbayev University, Astana, Kazakhstan}
\maketitle
\IEEEpeerreviewmaketitle
\begin{abstract}
The cryptographic security provided by various techniques of random number generator (RNG) construction is one of the developing researches areas today. Among various types of RNG, the true random bit generator (TRBG) can be considered as the most unpredictable and most secured because its randomness seed is generated from chaotic sources. This paper proposes a design of TRBG model based on double-scroll attractors circuits with GST memristor. After implementation and simulation of the chaotic circuit with GST memristor emulator, the chaotic behavior of the output voltage and inductor current were received. Moreover, their dependence on the input voltage revealed the close to double-scroll form. The randomness generated from the proposed circuit was tested by receiving Fast Fourier Transform (FFT) and Lyapunov exponents of the output voltage. 
\end{abstract}
\section{Introduction}
\IEEEPARstart{T}{he} Random Number Generator (RNG) is widely used subject in various technology spheres. Thus, it is one of the dominant tools in cryptographic security today and its importance gradually rises with the technological development. There are many different techniques to create RNGs based on their classification: true random number generator (TRNG), pseudo random number generator and hybrid number generator. As TRNG is generated from natural/non-deterministic and chaotic sources, its randomness seed can be considered as the most unpredictable and consequently most secured. One example of the non-deterministic source is the chaotic oscillator which generates double-scroll attractors \cite{IEEEhowto:yal}, \cite{IEEEhowto:cao}. This chaotic oscillator is modeled by analogy of the circuit implemented by Chua in 1971, which is essentially oscillator with non-linear resistor for executing chaotic behavior inside \cite{IEEEhowto:chua}.  Moreover, relatively new and full of potential memristive technology allows to replace non-linear resistor in the proposed chaotic circuit for TRNG because of its non-linear and distinctive electrical properties. Memristors are differed from each other and classified according to the technology of creation, materials and related features. In this paper we propose a design of TRNG circuit that is based on Phase- Change Memory device ($Ge_{2}$$Sb_{2}$$Te_{2}$), which is known as GST memristor \cite{IEEEhowto:wou}, \cite{IEEEhowto:yang}, \cite{irmanova2018neuron}, \cite{irmanova2017multi}. For the present time there is no existing GST memristor model, that is why its characteristical equations and parameters should be investigated in order to create its emulator circuit. GST memristor emulator circuit is based on the different resistance levels. According to Li et al., the emulator can be constructed from the three serial resistors with different values and parallel to them two capacitors. 
\\In order to reveal the effect of the addition of GST memristor emulator circuit, the theoretically and practically confirmed Chua's circuit with non-linear resistor implemented as two diodes with different polarities and two pairs of resistors with different values should be also constructed \cite{IEEEhowto:ken}, simulated and compared with studying circuit. 

This paper is organized as follows. We first investigate the properties of GST memristor and construct its emulator circuit in LTSpice software. Then, mathematical calculations and analysis for GST memristor and other devices parameters are performed. After that, we analyze and compare the original Chua’s circuit with two diodes and the modified circuit with GST memristor model in order to receive double-scroll attractors and reveal the most efficient one. 
\section{GST memristor emulator circuit}
\begin{figure}[H]
  \centering
  \includegraphics[width=0.4\textwidth,height=0.3\textheight,keepaspectratio]{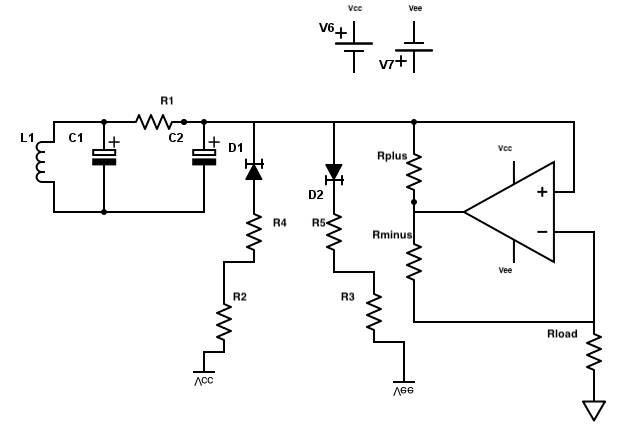}
  \caption{Chua circuit with two diodes and resistors implementing non-linear resistor behavior\cite{IEEEhowto:ken}}
  \label{f1}
\end{figure}

\begin{figure}[H]
  \centering
  \includegraphics[width=0.3\textwidth,height=0.2\textheight,keepaspectratio]{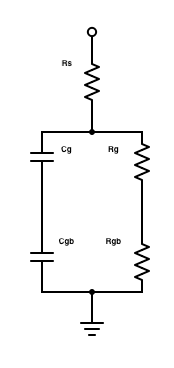}
  \caption{GST memristor emulator circuit}
  \label{f2}
\end{figure}

\begin{figure}[H]
  \centering
  \includegraphics[width=0.5\textwidth,height=0.4\textheight,keepaspectratio]{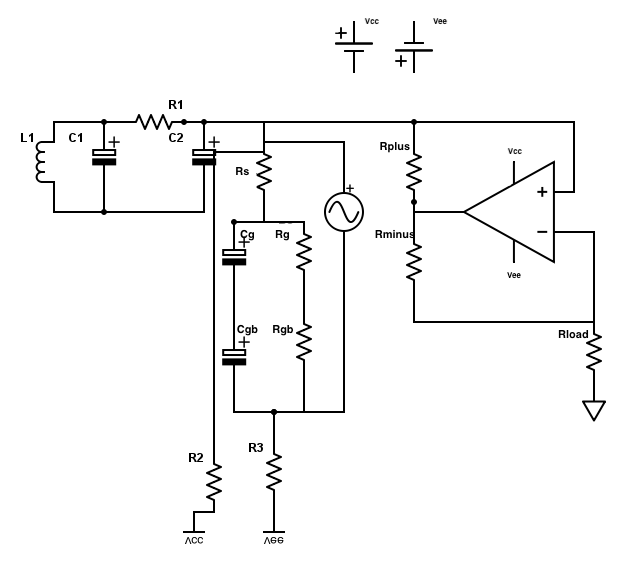}
  \caption{Chua circuit with implemented GST memristor emulator circuit}
  \label{f3}
\end{figure}

The chaotic circuit constructed in this paper is based on the fundamental Chua's circuit with non-linear resistor implementation. Initially, as it was proposed by Kennedy \cite{IEEEhowto:ken}, the original circuit with non-linear resistor, which is implemented with two diodes with different polarities and two pairs of resistors of 3.3 k$\Omega$ and 47k$\Omega$ values, and negative resistor, which consists of operational amplifier and three resistors of 290$\Omega$, 290$\Omega$ and 1.2k$\Omega$ values, is constructed. The designed original circuit is demonstrated on the Fig. \ref{f1}.
In order to implement new modified Chua's circuit with GST memristor, the GST memristor's emulator circuit with 3 different levels resistors and two capacitors was constructed because there is no already designed model of GST memristor. As it was suggested by Li et al. \cite{IEEEhowto:lit}, the emulator circuit consisting of contact resistor of the electrodes $R_{s}$ serially connected to parallel connection of pure resistor and capacitor of GST ($R_{p}$ and $C_{p}$). In order to better represent equivalent behavior of the GST memristor two main sources of defect in crystalline grain and at grain boundaries should be considered. That is why, $R_{p}$ and $C_{p}$ are separated to $R_{g}$ and $C_{g}$ and $R_{gb}$ and $C_{gb}$, that corresponds to the first and second types of defects. The corresponding circuit of the GST memristor emulator can be observed on the Fig. \ref{f2}.
Thus, replacement of the non-linear resistor from the original circuit with the GST memristor emulator shown on the Fig. \ref{f2} leads to the design of the new Chua's circuit with  GST memristor emulator, as it can be seen from the Fig. \ref{f3}. 
\section{Mathematical Analysis}
The mathematical analysis of the new chaotic circuit with the GST memristor emulator starts from revealing characteristic equations of the main device in the circuit - GST memristor emulator. The relationship between all components of the emulator circuit, which were described in the previous section, can be characterized by following equation for the total impedance of the  circuit:
\begin{equation}\label{one}
	Z= R_{s} + \frac{R_{g}}{1+jwR_{g}C_{g}}+\frac{R_{gb}}{1+jwR_{gb}C_{gb}}
\end{equation}
\\Based on the Eq. (\ref{one}) generally the memristor emulator circuit was constructed (Fig. \ref{f2}) and the corresponding values were selected, which will be discussed in the discussion section. Other important equations for characterization of the GST memristor were described by the Xiao et el. \cite{IEEEhowto:xiao}, the characteristic equations for the GST memristor are \cite{IEEEhowto:xx}: 
\begin{equation}\label{two}
	lI= M^{-1}V_{M}
\end{equation}
\begin{equation}\label{three}
	M= f(V_{M})[\theta(V_{M})\theta(\frac{M}{R_{1}}-1)+\theta(-V_{M})\theta(1-\frac{M}{R_{2}})]\times\gamma(1+W)
\end{equation}
\begin{equation}\label{four}
	f(V)= -{\beta}V + \frac{\beta-\alpha}{2}(|V+V_{L}|-|V-V_{R}|+V_{R}-V_{L})
\end{equation}
\begin{equation}\label{five}
	W= W\times\phi(W)(W-W_{t})
\end{equation}
Here, $V_{M}$ and I are voltage and current passing through the memristor, M and W represents memristance and phase of GST, and $V_{L}$ and $V_{R}$ are threshold voltages. $\theta$() is the unit step function which limits memristance to the $R_{1}$ and $R_{2}$ values. Also, $\alpha$ and $\beta$ are characteristic rates of change of memristor, depending whether $V_{M}$ is less or greater than threshold voltage, respectively. Moreover, $\gamma$ is the correction factor for the variation of memristance because of phase transition, while $W_{t}$ is the threshold of phase transition and $\phi$(W) is the mapping function of phase. The nature of these equations is still under the study.

Secondly, it is important to analyze the total chaotic circuit. After analysis of the chaotic circuit with HP memristor by Muthuswamy \cite{IEEEhowto:mut}, the following characteristic differential equations for the circuit components without considering memristor effects were derived by him: 
\begin{equation}\label{eiteen}
\frac{d\phi}{dt}=V_{1}(t)
\end{equation}
\begin{equation}\label{niteen}
\frac{dV_{1}}{dt}=\frac{1}{C_{1}}\times[\frac{V_{2}-V_{1}}{R_{1}} - i(t)]
\end{equation}
\begin{equation}\label{twenty}
\frac{dV_{2}}{dt}=\frac{1}{C_{2}}\times[\frac{V_{1}-V_{2}}{R_{1}} - i_{L}(t)]
\end{equation}
\begin{equation}\label{21}
\frac{di_{L}}{dt}=\frac{V_{2}}{L} 
\end{equation}
\\Taking into the account the memristor effect and using the above four equations the nonlinearity equation of the charge in chaotic circuit was derived and taken as cubic \cite{IEEEhowto:mut}:
\begin{equation}\label{22}
q=\theta\phi + \sigma\phi^{3}
\end{equation}
\\From the Eq. \ref{22} the next one for memductance can be derived using derivatives:
\begin{equation}\label{23}
W(\phi)=\frac{dq}{d\phi}=\theta+ \sigma\phi^{2}
\end{equation}
\\All the above equation can be related to our chaotic circuit with GST memristor, as the circuit is close to ours. All components from the HP memristor's emulator circuit can be generally compared with the components of GST memristor's emulator circuit.   Therefore,  $\theta$ and $\sigma$ values, which are constants from the Eq. \ref{22} and Eq. \ref{23}, for GST memristor emulator circuit can be characterized with the following equations:
\begin{equation}\label{24}
	\alpha = -\frac{1}{R_{load}}
\end{equation}
\begin{equation}\label{25}
	\beta = \frac{1}{3}(\frac{R_{g}+R_{gb}}{R_{gb}*R_{load}*R_{c}})
\end{equation}

\begin{figure}[H]
  \centering
  \includegraphics[width=0.5\textwidth,height=0.4\textheight,keepaspectratio]{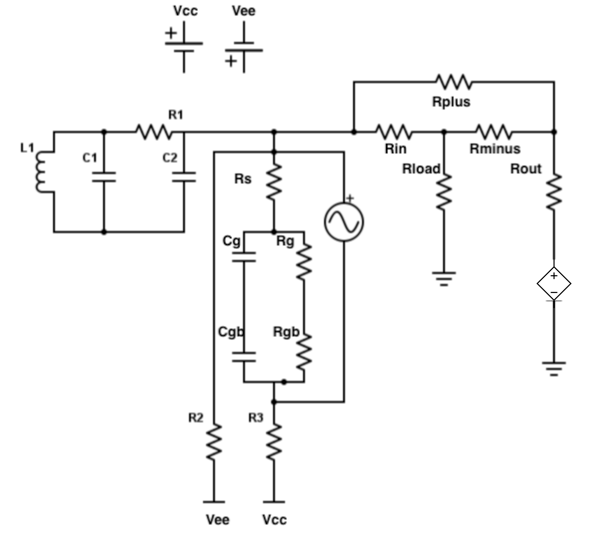}
  \caption{Chua circuit with implemented GST memristor emulator circuit and small-signal circuit  for operational amplifier in the Negative Resistor}
  \label{f4}
\end{figure}

\begin{figure}[H]
  \centering
  \includegraphics[width=0.3\textwidth,height=0.3\textheight,keepaspectratio]{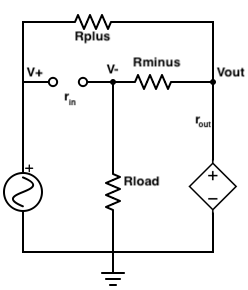}
  \caption{Small-signal model for  operational amplifier in Negative Resistor with open-circuited input resistance and short-circuited output resistance}
   \label{f5}
\end{figure}

\begin{figure}[H]
  \centering
  \includegraphics[width=0.35\textwidth,height=0.35\textheight,keepaspectratio]{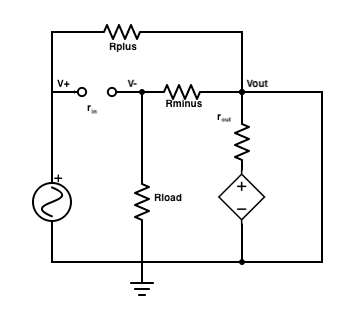}
  \caption{Small-signal model for operational amplifier in Negative Resistor with open-circuited input resistance}
  \label{f6}
\end{figure}

Another important part of the chaotic circuit that must be analyzed is the negative resistor. In order to mathematically analyze the negative resistor circuit, which is designed with operational amplifier and three resistors (see Fig. \ref{f3}), it is necessary to provide the small-signal analysis. Small-signal analysis is the method to express behavior of non-linear device in terms of linear equations. As it can be seen from the Fig. \ref{f4}, the operational amplifier was replaced by voltage controlled voltage source, input and output resistors, which is basically the small-signal model circuit.
Parametric equations for the circuit on the Fig. \ref{f4} can be derived by analyzing it in the following way.
Let the operational amplifier be considered as an ideal one. It means, that resistors $r_{in}$ and $r_{out}$, which are parameters inside operational amplifier, are equal to infinity and zero values. Thus, during the analysis $r_{in}$ is replaced by open- circuit notation and rout with short-circuit. In order to properly derive equations of small-signal circuit, it should be considered more specifically as on the Fig. \ref{f5}. First of all, the equation for resistance of the input source $R_{in}$ should be calculated.
From the nodal analysis of the circuit on the Fig. \ref{f5}, it is clear that voltage at the node V+ represented by Eq. (\ref{six}):
\begin{equation}\label{six}
	V+ = V_{out} − i*R_{plus} = V_{out} − V_{s}
\end{equation}
\\Where Vs is the voltage supplied by the source, in this case sinusoidal signal connected with memristor, and it is completely equal to the i${\times}$$R_{plus}$ because of infinitely large $r_{in}$. And $V_{out}$ is the output voltage of the operational-amplifier. On the other side of the open-circuit part the following equation at the node V- using Voltage divider principle can be derived:
\begin{equation}\label{seven}
	V-=V_{out}\frac{R_{minus}}{R_{minus} + R_{load}}
\end{equation}
Due to the open circuit between V+ and V-, they can be considered as equal. Equalization of Eq. (\ref{six}) and Eq. (\ref{seven}) leads to the following:
\begin{equation}\label{eight}
	V_{out} - V_{s} = V_{out}\frac{R_{minus}}{R_{minus} + R_{load}}
\end{equation}
Derivation of $V_{out}$ from the:
\begin{equation}\label{nine}
	V_{out} = V_{s}(1+\frac{R_{minus}}{R_{load}})
\end{equation}
In order to find input current i, the nodes V+ and $V_{out}$ should be considered. As $R_{plus}$ is the positive feedback resistance, input current goes from $V_{s}$ to $V_{out}$ through it:
\begin{equation}\label{ten}
	i= \frac{V_{s}-V_{out}}{R_{plus}}=\frac{V_{s}-V_{s}(1+\frac{R_{minus}}{R_{load}})}{R_{plus}}=\frac{-V_{s}*R_{minus}}{R_{plus}*R_{load}}
\end{equation}
Consequently, resistance of the input source $R_{in}$ is calculated
simply by division of $V_{s}$ to (\ref{ten}).
\begin{equation}\label{eleven}
	V_{out} = \frac{V_{s}}{\frac{-V_{s}*R_{minus}}{R_{plus}*R_{load}}} = -\frac{R_{plus}*R_{load}}{R_{minus}} 
\end{equation}
Hence, input resistance of the negative resistor circuit has negative value.
\\Secondly, the equation for the output resistance of the negative resistor circuit should be calculated. In order to find it out, open circuit output voltage and short circuit current are required. Open circuit voltage can be calculated by considering node $V_{out}$ with $r_{in}$ = $\infty$ and $r_{out}$ = 0 conditions and using voltage divider technique.
\begin{equation}\label{twelve}
V_{out}=A(V_{+} - V_{-})\frac{R_{load}+R_{minus}}{R_{load}+R_{minus}+r_{out}} 
\end{equation}
Inserting equations (\ref{six}) and (\ref{seven}) in (\ref{twelve}):
\begin{equation}\label{thteen}
V_{out}=A(V_{out}-V_{s}-V_{out}\frac{R_{minus}}{R_{minus} + R_{load}})\frac{R_{load}+R_{minus}}{R_{load}+R_{minus}+r_{out}} 
\end{equation}
After manipulations and solving for $V_{out}$ the following equation is received:
\begin{equation}\label{foteen}
V_{out}=\frac{AV_{s}(R_{load}+R_{minus})}{R_{load}(1-A)+R_{minus}+r_{out}}= V_{oc} 
\end{equation}
Then, in order to find short-circuited current $i_{sc}$ the circuit on the Fig. \ref{f6} has to be analyzed. 
According to the Fig. \ref{f6}, when the node $V_{out}$ is short-circuited, $V_{out}$=0, the V- also becomes 0, as it was derived in (\ref{seven}). Therefore, the following equation is received after KCL technique:
\begin{equation}\label{fiteen}
i_{sc}=\frac{AV_{s}}{r_{out}} + i =  \frac{AV_{s}}{r_{out}} + \frac{V_{s}}{R_{plus}} = V_{s}(\frac{A}{r_{out}} + \frac{1}{R_{plus}})
\end{equation}
Finally, the division of (\ref{foteen}) by (\ref{fiteen}) results in the following $R_{out}$ equation:
\begin{equation}\label{siteen}
R_{out}=\frac{V_{oc}}{i_{sc}}=\frac{AV_{s}(R_{load}+R_{minus})r_{out}R_{plus}}{(AR_{plus}+r_{out})(R_{load}(1-A)+R_{minus}+r_{out})}
\end{equation}
Considering the fact, that usually A$>$$>$1 (large numbers) equation (\ref{siteen}) can be approximated as:
\begin{equation}\label{seteen}
R_{out}=-\frac{(R_{load}+R_{minus})*r_{out}*R_{plus}}{A*R_{plus}*R_{load}}
\end{equation}
\section{Simulation results}
The simulation of this circuit with GST memristor emulator was done on the LTSpice software and the following results were received:
\begin{figure}[H]
  \centering
  \includegraphics[width=0.40\textwidth]{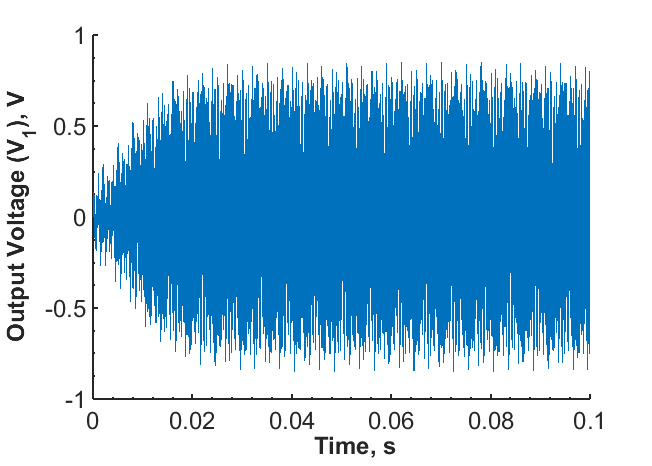}
  \caption{The Transient characteristic of output Voltage ($V_{1}$)}
  \label{f12}
\end{figure}
\begin{figure}[H]
  \centering
  \includegraphics[width=0.40\textwidth]{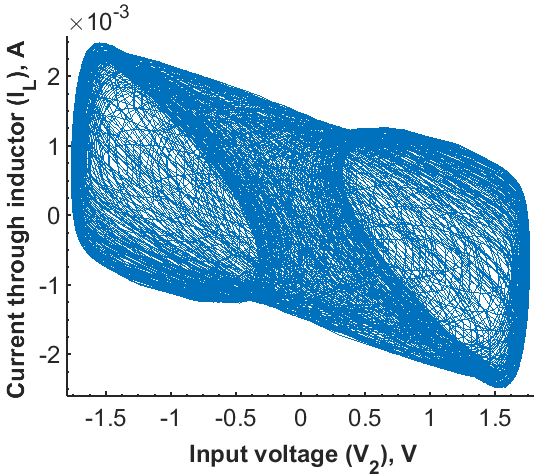}
  \caption{The current of inductor $I_{L}$ and input voltage $V_{2}$ characteristic}
  \label{f13}
\end{figure}
\begin{figure}[H]
  \centering
  \includegraphics[width=0.50\textwidth]{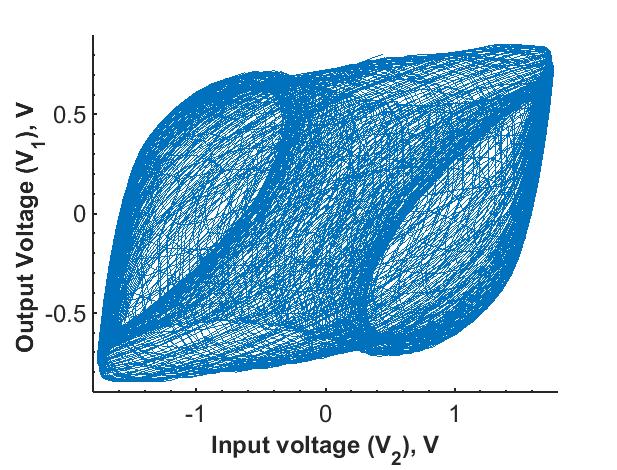}
  \caption{The output voltage $V_{1}$ and input voltage $V_{2}$ characteristic}
  \label{f14}
\end{figure}
\begin{figure}[H]
  \centering
  \includegraphics[width=0.40\textwidth]{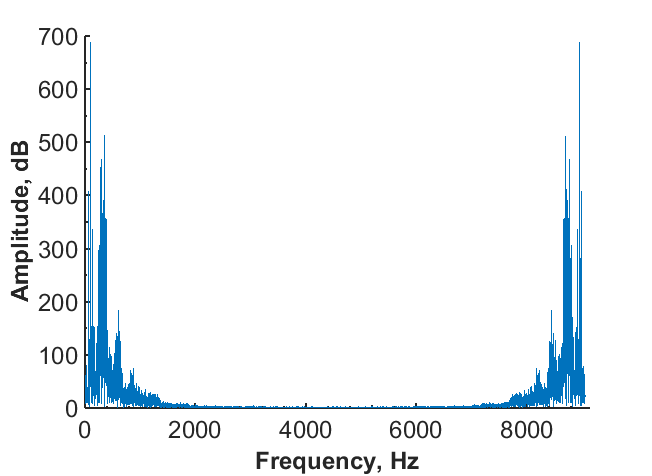}
  \caption{The Fast Fourier Transform of output voltage $V_{2}$}
  \label{f15}
\end{figure}
\begin{figure}[H]
  \centering
  \includegraphics[width=0.40\textwidth]{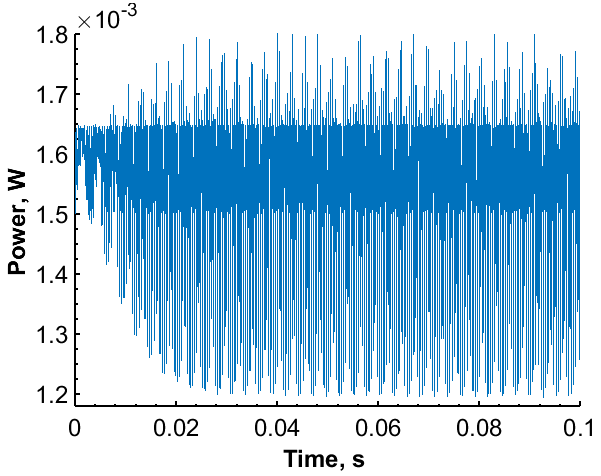}
  \caption{Power of the chaotic circuit with GST memristor emulator circuit implemented with resistors and capacitors}
  \label{f16}
\end{figure}

The constructed with GST memristor emulator Chua circuit has visually chaotic behavior for the voltage transient characteristic, as it can be seen from the Fig. \ref{f12}. In order to demonstrate the randomness of the signal spectra, the FFT of output voltage was performed, as it can be seen on the Fig. \ref{f15}. The significant variations in amplitude corresponding to different frequency values prove
the randomness of the signal. However, the relationship of inductor current with input voltage and input-output voltages relationship plots do not have perfectly corresponding to theory double-scroll forms, but generally resembling double scrolls with chaotic relations with each other (Figs. \ref{f13}-\ref{f14}). The power across the each component of GST memristor emulator circuit was found from the simulation and its power/time dependence across op-amp is shown on the Fig. \ref{f16}. 
\section{Discussion}

In this paper we demonstrate the simulation for the Chua's original chaotic circuit with non-linear resistor consisting of diodes and resistors together with modified circuit with GST memristor emulator circuit. This section aims to compare results of both circuits in order to reveal the effect of the memristor addition to the circuit.
\subsection{Fast Fourier Transform}

First of all, as it can be concluded from  Fig. \ref{f15}, where FFT of output voltage signals is illustrated, generally the output signal can be counted as random. As it was revealed during the research, the peak amplitude values for the original circuit attain approximately 600, while for the memristive circuit it reaches 700. Hence, deviations in voltage values for original circuit smaller than for circuit with memristor emulator, therefore the circuit with GST memristor emulator can be counted as better in terms of chaos.
\subsection{Power analysis}

Secondly, the general comparison of both circuits powers can be done in the following way. The original circuit requires 9V value for $V_{cc}$ and $V_{ee}$ for generating best chaotic output, while modified circuit  requires only 3V. It means that second circuit has approximately 3 times less power dissipation. However, the real analysis revealed the following result. According to average calculation of data from each component in non-linear circuit of original chaotic Chua circuit the power is 5.02 mW. While the calculation of average power of each component in GST memristor emulator circuit of new chaotic Chua circuit gave 4.47 mW. It can be  concluded, that the less power consumption of the Chua circuit with GST emulator is proved and the difference between it and the original circuit's power is 0.55 mW. Also, from the Fig. \ref{f16} it can be visually seen that power in the circuit with GST memristor varies from about 1.20 mW to 1.80 mW, whereas in original circuit it varies from approximately 4 mW to 26 mW. In spite of receiving small variations in values, low power consumption of the circuit with GST memristor emulator is still big advantage. 
\subsection{Area analysis}

Thirdly, another important comparison parameter is the area calculation. In order to calculate area of the original chaotic circuit with non-linear resistor circuit, the standard values of areas for diodes and resistors are taken. Diode 1N4148 dimensions are: 3.4${\times}$1.75 $mm^{2}$ for the cathode and 25.4x0.55 $mm^{2}$ for wires. Regular chip resistor dimensions are 0.6${\times}$0.3 $mm^{2}$. In total, the are of about 162 $mm^{2}$ will be occupied by the Chua circuit with non-linear resistor consisting of  two different polarities diodes and resistors. For the chaotic circuit with GST memristor, assuming that the designed GST memristor model will be used, the total occupied area is assumed to be about 1x1 $um^{2}$, where the GST layer thickness is 150 nm, and for the electrodes is 100 nm, according to Li et al \cite{IEEEhowto:lit}. Hence, the area for the GST memristor model 162000 times smaller than the area for the non-linear resistor. If the emulator circuit is considered, standard dimensions for the chip capacitor are 0.51${\times}$0.25 $mm^{2}$. The overall area occupied by GST emulator circuit is approximately 1.155 $mm^{2}$, which is still 140 times lesser than the area of the non-linear resistor. Thus, in terms of area parameter it is proved that the advantage on the chaotic circuit with GST memristor side.
\subsection{Lyapunov exponents randomness test}

\begin{figure}
  \centering
  \includegraphics[width=0.40\textwidth]{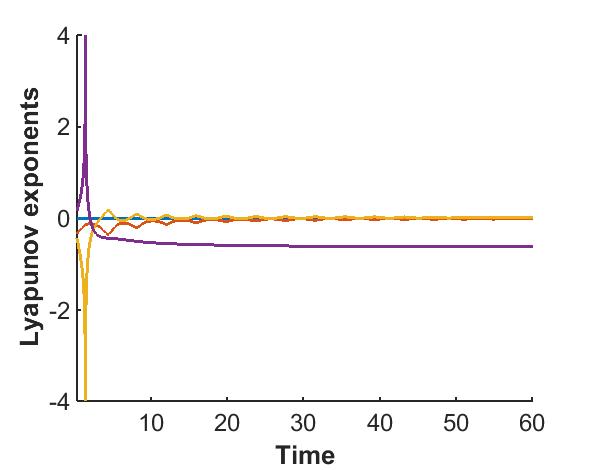}
  \caption{Lyapunov Exponents from the chaotic circuit with GST memristor emulator with corresponding $R_{g}$ = 25.9k$\Omega$. $C_{g}$=30.6pF, $R_{gb}$ = 280$\Omega$ and $C_{gb}$=5pF parameters}
   \label{f17}
\end{figure}

Moreover, in order to better understand randomness in the constructed circuit, the Lyapunov exponents were generated in MatLab software. Corresponding $\theta$ and $\sigma$ values for the characteristic memductance equation of the memristor circuit calculated by Eq. \ref{24} and Eq. \ref{25}, are 0.4${\times}$$10^{-3}$ and 1.35${\times}$$10^{-6}$ respectively. The Lyapunov exponents for the received $\sigma$ and $\theta$ parameters are shown on the Fig. \ref{f17}.
As it can be seen from the Fig. \ref{f17}, there are two exponents that have similar pic values 4 but with different signs at the beginning. After few time firstly positive exponent (purple color) becomes negative and stable, while firstly negative exponent (yellow color) becomes fluctuating around x axis. Generally, both exponents are in close agreement and their sum (red color) is fluctuating negative exponent, which proves presence of the chaos in the circuit \cite{IEEEhowto:mut}.  
\subsection{Values of components}

\begin{figure}[H]
  \centering
  \includegraphics[width=0.5\textwidth]{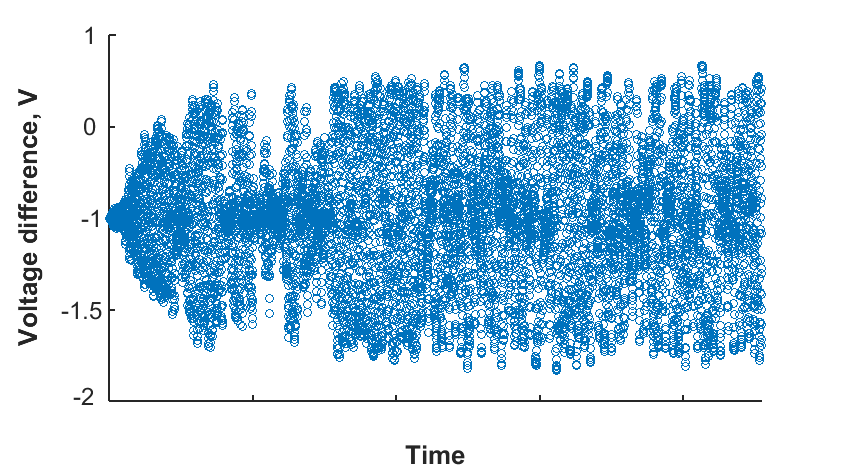}
  \caption{Difference of voltages corresponding to $R_{g}$=25.9k$\Omega$ and $R_{g}$=40.9k$\Omega$}
\label{f18}
\end{figure}

Finally, the values of used parameters for the GST memristor emulator circuit are $R_{s}$=100$\Omega$, $R_{g}$=25.9k$\Omega$, $R_{gb}$=280$\Omega$, as it is recommended in Li et al. \cite{IEEEhowto:lit}, and capacitors values as $C_{g}$=5 pF and $C_{gb}$=30.6 pF are selected, as it is recommended by Li et al.\cite{IEEEhowto:li}.  Parameters for the rest circuit general elements are: $R_{1}$=2k$\Omega$, $C_{2}$=100nF, $C_{1}$=10nF, $L_{1}$=18mH, $R_{plus}$=250$\Omega$, $R_{minus}$=230$\Omega$ and $R_{load}$=2.5k$\Omega$. Also, sinusoidal voltage source with 1kHz frequency and 11V amplitude is connected to memristor emulator circuit part. The variations of general elements in the circuit significantly influence on the output. That is why, these parameters are most appropriate to be used. On the same time, variations of GST memristor emulator parameters do not provide visual differences in the output. However, when the $R_{g}$ value is changed from 25.9k$\Omega$ to 40.9k$\Omega$, the corresponding output data were collected and their difference was observed and plotted. As it can be observed from the Fig. \ref{f18}, difference between voltage values for different resistances is also chaotic and achieves the maximum of approximately 2 times and minimum of close to 0. Thus, as the change is highly variable, it is difficult to observe it in the output of the chaotic circuit.   
\section{Conclusion}
This paper aimed to show that chaotic behavior can be achieved from the Chua's oscillator circuit with GST memristor emulator. After construction and simulation on LTSpice software of the original Chua's circuit with two diodes of different polarities and two pairs of different level resistors, which implements non-linear resistor functions, and of the modified Chua's circuit with GST memristor emulator circuit results were received and compared. The chaotic output from the targeted memristive circuit was obtained, which is the expected result. Deviations in amplitude of randomness according to FFT for the new circuit to about 100 greater than in original circuit. Also, comparison in terms of area and power indicates the advantage of the Chua circuit with GST memristor emulator circuit. Additionally, Lyapunov exponents randomness test proved the presence of the chaos in the new Chua's circuit with GST memristor emulator. More investigations of memristor emulator circuit and its parameters analysis are needed. It is recommended to study the addition of GST memristor emulator circuit to the chaotic circuits which generates N-scroll attractors, in order to receive more chaotic behavior.


\begin{thebibliography}{1}
\bibitem{IEEEhowto:yal}
M.~Yalçın, J.~Suykens and J.~Vandewalle, \emph{True Random Bit Generation From a Double-Scroll Attractor},  IEEE TRANSACTIONS ON CIRCUITS AND SYSTEMS-I: REGULAR PAPERS,vol.~51, num.~7.\hskip 1em plus
  0.5em minus 0.4em\relax IEEE, July  2004 [Online]. DOI: 10.1109/TCSI.2004.830683 
  
\bibitem{IEEEhowto:cao}
F.~Cao and F.~Li, \emph{ A Double-Scroll Based True Random Number Generator with Power and Throughput Adjustable}, ASICON '09, IEEE 8th International Conference on ASIC, Changsha, Hunan, China, October 20-23, 2009.\hskip 1em plus
  0.5em minus 0.4em\relax IEEE, October  2009 [Online]. pp. 309-312. DOI: 10.1109/ASICON.2009.5351446 
  
\bibitem{IEEEhowto:chua}
L.~Chua, \emph{Memristor-The missing circuit element},   IEEE Transactions on Circuit Theory, vol.~18, num.~5, pp. ~507-519\hskip 1em plus
  0.5em minus 0.4em\relax IEEE, September  1971 [Online]. DOI: 10.1109/TCT.1971.1083337
   
\bibitem{IEEEhowto:wou}
D.~J.~Wouters, R.~Waser and M.~Wutting, \emph{Phase-Change and Redox-Based Resistive Switching Memories},  Proceedings of the IEEE, vol.~103, num.~8, pp.~1274-1288.\hskip 1em plus
  0.5em minus 0.4em\relax IEEE, August  2015 [Online]. DOI: 10.1109/JPROC.2015.2433311

\bibitem{IEEEhowto:yang}
C.~Yang, B.~Liu, Y.~Wang, Y.~Chen, H.~Li ,  H.~Zhang   and G.~Sun, \emph{The Applications of NVM Technology in Hardware Security}, Proc. Great Lakes Symp. VLSI (GLVLSI), Boston, MA, USA, May 18-20, 2016. \hskip 1em plus
  0.5em minus 0.4em\relax IEEE, August 2016 [Online], pp. 311-316. DOI:10.1145/2902961.2903043
  


\bibitem{irmanova2018neuron}
A. Irmanova and A. P. James,  \emph{Neuron inspired data encoding memristive multi-level memory cell}, Analog Integrated Circuits and Signal Processing, Springer, p.1-6, \hskip 1em plus
0.5em minus 0.4em\relax 2018.

  \bibitem{irmanova2017multi}
A. Irmanova and A. P. James,  \emph{Multi-level Memristive Memory with Resistive Networks}, arXiv preprint arXiv:1709.04149,  \hskip 1em plus
0.5em minus 0.4em\relax 2017.
  
\bibitem{IEEEhowto:ken}
M.~P.~Kennedy, \emph{Three Steps to Chaos-Part
11: A Chua’s Circuit Primer},  IEEE TRANSACTIONS ON CIRCUITS AND SYSTEMS-I: FUNDAMENTAL THEORY AND APPLICATIONS, vol.~40, num.~10, pp.~657-674.\hskip 1em plus
  0.5em minus 0.4em\relax IEEE, October  1993 [Online]. DOI: 10.1109/81.246141

\bibitem{IEEEhowto:lit}
Y.~Li, Y.~Zhong, J.~Zhang, L.~Xu, Q.~Wang, Q.~Xu, H.~Sun and X.~Miao, \emph{Intrinsic memristance mechanism of crystalline stoichiometric Ge2Sb2Te5}, Applied Physics Letters, vol.~103, num.~4.\hskip 1em plus
  0.5em minus 0.4em\relax AIP, July 2013. DOI: 10.1063/1.4816283

\bibitem{IEEEhowto:xiao}
S.~Xiao, X.~Xie, S.~Wen, Z.~Zeng, T.~Huang  and J.~Jiang, \emph{GST-memristor-based online learning neural networks}, Neurocomputing, vol.~272, pp.~677-682.\hskip 1em plus
  0.5em minus 0.4em\relax Elsevier B.V., August 2017 [Online]. DOI:10.1016/j.neucom.2017.08.014 
  
  
      \bibitem{IEEEhowto:xx}
  Alex James, Memristor and Memristive Neural Networks, Intech, 2018, DOI: 10.5772/66539; ISBN: 978-953-51-3948-5
  
  
  
\bibitem{IEEEhowto:li}
Y.~Li, Y.~Zhong, J.~Zhang, L.~Xu, Q.~Wang, H.~Sun and J.~Jiang, \emph{Ultrafast synaptic events in a chalcogenide memristor}, Scientific reports, vol.~3, p.~1619.\hskip 1em plus
  0.5em minus 0.4em\relax Nature Publishing Group, April 2013.

\bibitem{IEEEhowto:mut}
B.~Muthuswamy, \emph{Implementing Memristor Based Chaotic
Circuits}, International Journal of Bifurcation and Chaos, vol.~20, num.~5, pp.~1335-1350.\hskip 1em plus
  0.5em minus 0.4em\relax World Scientific Publishing Company, January  2010 [Online]. DOI: 10.1142/S0218127410026514
  
  

\end{thebibliography}
\end{document}